# Machine Learning-Based Detection of Parkinson's Disease From Resting-State EEG: A Multi-Center Study

Anna Kurbatskaya[1], Alberto Jaramillo-Jimenez[2,3,4]
John Fredy Ochoa-Gomez[5], Kolbjørn Brønnick[2,3] and Alvaro Fernandez-Quilez[1,4]

*Abstract*— Resting-state EEG (rs-EEG) has been demonstrated to aid in Parkinson's disease (PD) diagnosis. In particular, the power spectral density (PSD) of low-frequency bands ($\delta$ and $\theta$) and high-frequency bands ($\alpha$ and $\beta$) has been shown to be significantly different in patients with PD as compared to subjects without PD (non-PD). However, rs-EEG feature extraction and the interpretation thereof can be time-intensive and prone to examiner variability. Machine learning (ML) has the potential to automatize the analysis of rs-EEG recordings and provides a supportive tool for clinicians to ease their workload. In this work, we use rs-EEG recordings of 84 PD and 85 non-PD subjects pooled from four datasets obtained at different centers. We propose an end-to-end pipeline consisting of preprocessing, extraction of PSD features from clinically-validated frequency bands, and feature selection before evaluating the classification ability of the features via ML algorithms to stratify between PD and non-PD subjects. Further, we evaluate the effect of feature harmonization, given the multi-center nature of the datasets. Our *validation* results show, on average, an improvement in PD detection ability (69.6% vs. 75.5% accuracy) by logistic regression when harmonizing the features and performing univariate feature selection ($k = 202$ features). Our final results show an average global accuracy of 72.2% with balanced accuracy results for all the centers included in the study: 60.6%, 68.7%, 77.7%, and 82.2%, respectively.

*Clinical relevance*— We present an end-to-end pipeline to extract clinically relevant features from rs-EEG recordings that can facilitate the analysis and detection of PD. Further, we provide an ML system that shows a good performance in detecting PD, even in the presence of centers with different acquisition protocols. Our results show the relevance of harmonizing features and provide a good starting point for future studies focusing on rs-EEG analysis and multi-center data.

## I. INTRODUCTION

Parkinson's disease (PD) is clinically characterized by both motor symptoms, such as bradykinesia, rigidity, and rest tremor, and non-motor symptoms, such as cognitive deficits [1]. Pathologically, the primary features of PD are the loss of nigrostriatal dopaminergic neurons caused by $\alpha$-synuclein aggregates [2]. Accurate diagnosis at the early stages of PD increases the chance of success for neuroprotective disease-modifying treatments. However, the diagnostic accuracy based on clinical examination is not satisfactory [3], especially in the early stages of the disease, when the response to dopaminergic treatment is low [4]. On the other hand, the non-invasive measurement of $\alpha$-synuclein aggregates by positron emission tomography (PET) is not available yet, while the quantitative measurement of nigrostriatal dopaminergic neurons by single photon emission computed tomography (SPECT) is expensive and limited due to radiation dose [5].

Quantitative electroencephalography (qEEG) serves as a convenient, inexpensive, and non-invasive technique for analyzing the brain's electrical activity. Some qEEG features have been shown to be promising biomarkers for PD [6]. For instance, it has been demonstrated that increased power spectral density (PSD) of $\delta$ and $\vartheta$ bands and decreased PSD of $\alpha$ and $\beta$ bands are typically connected to the progression of PD [7], [8]. To extract essential features from EEG signals, advanced signal-processing techniques must be applied. Moreover, the application of automatized diagnostic systems holds the potential to overcome the limitations of visual analysis of EEG [9], which remains the gold standard for clinical interpretation.

Over the last few years, several studies have aimed to combine qEEG and machine learning (ML) techniques for the automated detection of PD. Several ML methods have been tested in [10], using the relative power, median, and peak frequencies in $\vartheta$, $\delta$, $\alpha$, and $\beta$ frequency bands as features in addition to the $\alpha/\vartheta$ ratio for differentiation between subjects with PD and without PD (non-PD). In addition, penalized logistic regression (LR) using LASSO was used to obtain a subset of best-performing features. In [11], an application of high-order spectra features for an automated diagnosis of PD was also performed, in which feature selection using the *t* value ranked the obtained features, and highly relevant features were used as input to several classifiers.

There are several pitfalls in EEG studies applying ML [12]. Most of the recently published studies do not clearly state whether subject-wise or record-wise classification was performed. Moreover, there are usually overly-optimistic performance metrics reported due to cross-contamination of training and test data [13]. Furthermore, many studies have re-used one of the previously published datasets for developing new models, e.g., the same datasets have been circulating and utilized by different research groups. The

*This work was not supported by any organization

[1]Anna Kurbatskaya and Alvaro Fernandez-Quilez are with the Department of Electrical Engineering and Computer Science, Faculty of Science and Technology, University of Stavanger, Stavanger, Norway. e-mails: anna.kurbatskaya@uis.no and alvaro.f.quilez@uis.no

[2,3] Alberto Jaramillo-Jimenez and Kolbjørn Brønnick are with the Centre for Age-Related Medicine (SESAM), Department of Psychiatry, Stavanger University Hospital, and the Department of Public Health, Faculty of Health Sciences, University of Stavanger, Stavanger, Norway. e-mails: alberto.jaramilloj@udea.edu.co and kolbjorn.s.bronnick@uis.no

[4] Alberto Jaramillo-Jimenez is with Grupo de Neurociencias de Antioquia, Universidad de Antioquia, School of Medicine, Medellín, Colombia.

[5] John Fredy Ochoa-Gomez is with Grupo Neuropsicología y Conducta, Universidad de Antioquia, School of Medicine, Medellín, Colombia. e-mail: john.ochoa@udea.edu.co

[6] Alvaro Fernandez-Quilez is with the Stavanger Medical Imaging Laboratory (SMIL), Department of Radiology, Stavanger University Hospital, Stavanger, Norway.

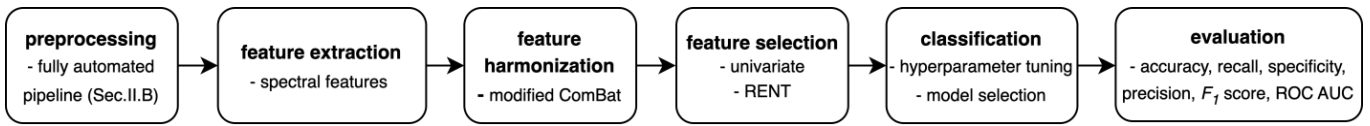

Fig. 1: End-to-end pipeline consisting of preprocessing, feature extraction, feature harmonization, feature selection, classification, and evaluation.

repeated use of the same datasets without external evaluation (by using other datasets) limits the replicability of the results. At the same time, employing different datasets, which are acquired by various devices and at different research centers using individual populations, improves generalization and provides more reliable performance. To the best of our knowledge, only in two of the recent studies two different datasets were used at the same time [14], [15].

Nevertheless, combining EEG recordings from several datasets has been reported to lead to increased Type I error for group-level analysis even with the use of a standard preprocessing pipeline for all recordings [16], [17]. To overcome the limitations of multi-center data pooling, solutions implemented in the fields of genetics and brain imaging have been widely applied to statistically model the features extracted from the raw signals [18], [19]. However, there are scarce implementations of harmonization methods in the EEG analysis [17], [20].

In this study, we aim to classify PD patients and non-PD subjects by applying ML techniques using resting-state EEG (rs-EEG) recordings obtained at different research centers. We present an end-to-end pipeline (Fig. 1) consisting of preprocessing, feature extraction, feature harmonization, feature selection, classification, and evaluation based on validation results (code is available on GitHub). We also examine how the harmonization of data and different feature selection techniques impacted the performance when classifying the subjects from all the centers together (globally) and each center separately.

## II. MATERIALS AND METHODS

### A. Data

In total, rs-EEG recordings of 169 subjects (84 PD and 85 non-PD) were included in this study. The recordings were obtained from four cross-sectional studies provided by four different centers in three countries: Colombia (Medellín), Finland (Turku), and the USA (Iowa City and San Diego). Some of the relevant study details for each dataset are given in Table I.

TABLE I: Relevant study details for each included dataset.

|  | Subjects | Diagnostic Criterion | Match |
|---|---|---|---|
| *Iowa City* [14] | 14 PD, 14 non-PD | UK Brain Bank [21] | age, gender, comparable education |
| *Medellín* [22] | 36 PD, 36 non-PD | MDS [23] | age, gender, years of education |
| *San Diego* [24] | 15 PD, 16 non-PD | not stated | age, gender, cognitive performance, handedness |
| *Turku* [25] | 19 PD, 19 non-PD | UK Brain Bank [21] and MDS [23] | age |

EEG recordings were acquired with eyes closed for the Medellín, San Diego, and Turku datasets and with eyes open for the Iowa City dataset. In addition, all the PD subjects of the Iowa City, Medellín, and San Diego datasets were in the ON phase of levodopa treatment during the recordings. For the Turku dataset, 13 PD subjects were assessed in the OFF phase, and the remaining subjects were in the ON phase. There were 29 common channels within all the datasets that were included in the analysis: AF3, AF4, C3, C4, CP1, CP2, CP5, CP6, Cz, F3, F4, F7, F8, FC1, FC2, FC5, FC6, Fp1, Fp2, Fz, O1, O2, Oz, P3, P4, P7, P8, T7, and T8.

### B. EEG Preprocessing

All the recordings were converted to the brain image data structure (BIDS) [26] and preprocessed using a Python-implemented pipeline, previously validated by [27]. The pipeline consisted of robust average re-referencing and adaptive line-noise correlation combined with detection and interpolation of noisy channels by the Python implemented [28] PREP pipeline, originally MATLAB based [29]. The next steps of the pipeline were high-pass finite impulse response (FIR) filtering at 1 Hz to eliminate low-frequency drifts and wavelet-based independent component analysis (ICA) to suppress muscular, eye-blinking, and eye-movement artifacts [30]. Following, the signals were low-pass FIR filtered at 30 Hz and segmented into epochs of five-second length. Epochs that were still compromised by artifacts were rejected by a procedure based on signal parameters, such as extreme amplitude and spectral power values, and statistical features, such as linear trend, joint probability, and kurtosis. The final number of epochs included in the analysis was determined based on the common number of epochs across all subjects within each dataset. The number of available epochs was 17, 44, 28, and 19 for the Iowa City, Medellín, San Diego, and Turku datasets, respectively.

### C. Classification Framework

*1) Feature Extraction:* For feature extraction, the frequency spectrum was divided into four bands recommended for clinical research: $\delta$ [1–4 Hz), $\vartheta$ [4–8 Hz), $\alpha$ [8–13 Hz), and $\beta$ [13–30 Hz) [6]. In addition, the $\vartheta$ band was divided into slow-$\vartheta$ [4–5.5 Hz) and fast-$\vartheta$, otherwise called "pre-$\alpha$", sub-bands [5.5–8 Hz) [31]. For each epoch, PSD values were obtained by dividing the PSD within a frequency (sub-)band by the total signal power in the 1-30 Hz range. The PSD values were computed by using the multi-taper method [32]. Moreover, we computed the relative $\alpha/\vartheta$ PSD ratio values using the relative PSD $\alpha$ and $\vartheta$ values [33]. Finally, we averaged the PSD values of all the epochs resulting in one value representing each spectral feature for a given channel [34], [35]. In total, it produced 203 features (7 features × 29 channels) per subject.

*2) Feature Harmonization:* All the averaged relative PSD values were log-transformed following the conventional procedure for spectral analysis in fixed frequency bands.

TABLE II: Global validation performance (mean [95% CI]) without feature selection.

| | Model | Accuracy,% | Recall,% | Specificity,% | Precision,% | $F_1$,% | ROC AUC |
|---|---|---|---|---|---|---|---|
| non-harmonized | LR | 69.6 [55.5 83.7] | 64.7 [43.0 86.4] | 74.9 [53.3 96.4] | 72.9 [56.3 89.5] | 67.3 [49.9 84.7] | 0.70 [0.56 0.84] |
| | SVM | 65.4 [50.9 79.8] | 58.2 [31.5 84.9] | 72.9 [57.7 88.0] | 67.6 [55.2 80.0] | 61.2 [42.0 80.3] | 0.66 [0.51 0.80] |
| harmonized | LR | 73.0 [59.2 86.9] | 69.7 [45.8 93.6] | 76.4 [69.0 83.7] | 73.1 [60.6 85.6] | 70.7 [51.4 90.1] | 0.73 [59.2 86.9] |
| | SVM | 73.0 [60.7 85.3] | 69.7 [56.1 83.3] | 76.2 [58.3 94.1] | 75.8 [62.3 89.2] | 72.1 [59.8 84.4] | 0.73 [0.61 0.85] |

TABLE III: Global validation performance (mean [95% CI]) with feature selection. $m$ is the number of top features to select for univariate feature selection; $\tau$ is the selection criterion for RENT.

| | Model | feature selection | Accuracy,% | Recall,% | Specificity,% | Precision,% | $F_1$,% | ROC AUC |
|---|---|---|---|---|---|---|---|---|
| non-harmonized | LR | univariate ($m$ = 68) | 73.7 [65.3 82.2] | 77.9 [72.3 83.5] | 69.2 [52.8 85.7] | 73.2 [61.8 84.6] | 75.1 [69.0 81.2] | 0.74 [0.65 0.82] |
| | | RENT ($\tau$ = 0.26) | 72.2 [58.0 86.4] | 70.0 [41.8 98.2] | 75.0 [58.9 91.0] | 74.0 [54.2 93.7] | 70.0 [50.8 89.3] | 0.73 [58.0 87.0] |
| | SVM | univariate ($m$ = 65) | 73.8 [66.2 81.3] | 72.9 [56.4 89.4] | 74.8 [57.8 91.9] | 76.3 [64.5 88.1] | 73.2 [64.7 81.6] | 0.74 [0.67 0.81] |
| | | RENT ($\tau$ = 0.12) | 70.4 [57.0 83.9] | 63.2 [35.2 91.2] | 78.3 [62.8 93.8] | 76.6 [58.4 94.8] | 66.2 [47.3 85.2] | 0.71 [0.57 0.84] |
| harmonized | LR | univariate ($m$ = 202) | **75.5 [66.3 84.7]** | 73.0 [56.7 89.3] | **78.0 [70.0 86.1]** | 76.7 [68.9 84.4] | 74.3 [62.6 86.0] | **0.76 [0.66 0.85]** |
| | | RENT ($\tau$ = 1.0) | 69.7 [48.4 91.0] | 66.4 [34.5 98.2] | 73.0 [61.3 84.7] | 67.4 [42.1 92.7] | 66.5 [37.3 95.8] | 0.70 [0.48 0.91] |
| | SVM | univariate ($m$ = 193) | 75.5 [65.4 85.5] | 69.7 [53.2 86.2] | 81.4 [71.3 91.5] | 79.1 [70.2 87.9] | 73.5 [61.4 85.7] | 0.76 [0.66 0.85] |
| | | RENT ($\tau$ = 1.0) | 69.7 [52.8 86.5] | 63.0 [33.6 92.4] | 76.2 [58.3 94.1] | 70.8 [52.1 89.4] | 65.1 [38.4 91.8] | 0.70 [0.53 0.86] |

These log-transformed features are further referred to as non-harmonized features. Subsequently, the log-transformed features were harmonized by using the modified ComBat harmonization model [18] with 1000 bootstrap repetitions, including gender, age, and diagnosis (PD and non-PD) as covariates. The Medellín dataset was used as the reference center as it was the largest individual dataset. The "batch" variable included four levels corresponding to each of the datasets.

*3) Classifiers:* To perform a *subject*-wise classification, several classifiers, including LR, k-nearest neighbors (k-NN) algorithm, decision tree (DT), and support vector machine (SVM), were trained and compared in terms of their validation accuracies.

To validate, train, and test the algorithms, the data were divided into training/test subsets with a 70/30 split stratified by both center and diagnosis attributes. The features belonging to one subject were a part of only either the training or test subset to avoid cross-contamination. Next, the training subset was used for hyperparameter tuning and model selection using nested cross-validation (CV) consisting of two 5-fold CV loops. Hyperparameter tuning was performed within the inner loop of the nested CV using a grid search approach over the whole search space. The list of hyperparameters and the values thereof can be found in the provided code. The outer loop estimated the generalization performance of the resulting classifiers.

*4) Performance Evaluation:* We evaluated the performance of the classifiers in terms of accuracy, recall, specificity, precision, $F_1$ score, and area under the receiver operating characteristic (ROC) curve (AUC) for both validation and testing. The best classifier was selected based on the average global validation accuracy, i.e., for all the centers together. Along with the mean values for each metric, we report 95% confidence intervals (CI) to define the precision of the estimate [36]. To obtain the final test performance metrics and estimate the uncertainty of the performance, the test subset was bootstrapped with replacement and the same size as the original test partition using 100 iterations.

*5) Feature Selection:* Given the high dimensional nature of the data (203 features per subject), we separately tested two feature selection methods to determine a subset of the features that positively impacted validation accuracy.

The first one is univariate feature selection based on ANOVA $F$-value with the number of top features to select as a parameter. The second one is the Repeated Elastic Net Technique (RENT), which relies on an ensemble of generalized linear models and evaluates the weight distributions of features across the models using three criteria as parameters [37]. Feature selection was performed inside the outer loop of the nested CV for each fold to reduce the bias through overfitting [38], and the following hyperparameter tuning was performed using the selected features within the inner loop. Both approaches for feature selection were repeated for different values of the parameters. Since different sets of features were defined by each iteration, we performed a merging of the selected along with the removal of duplicates. The final predictive model based on the validation performance was tested using the test subset with the remaining features.

III. RESULTS AND DISCUSSION

The global validation performance (mean [95% CI]) of the two best-performing classifiers, LR and SVM, without feature selection, are presented in Table II. Both models achieved better performance in terms of all the metrics using harmonized features compared to non-harmonized ones.

Table III shows the classification results using feature selection. It can be seen that the univariate feature selection improved the performance in terms of all the metrics for both LR and SVM using both non-harmonized and harmonized features in comparison with no feature selection, with the sole exception of specificity for LR (74.9% vs. 69.2%). On the other hand, RENT had a tendency to improve classification results for non-harmonized features but at the same time to worsen the results for harmonized ones.

In general, global validation improvement did not always imply good performance for every individual center. Both harmonization and feature selection gave different center-wise results. Even if the global accuracy shows promising results for harmonization, datasets from more centers need to be included in the study to validate the final results. Thus, if new centers want to implement the proposed pipeline in their studies, other approaches should be considered to ensure good individual performance for their specific population.

TABLE IV: Test results (mean [95% CI]) achieved with LR trained on harmonized features with univariate feature selection.

| | Accuracy, % | Recall, % | Specificity, % | Precision, % | $F_1$, % | AUC |
|---|---|---|---|---|---|---|
| global | **72.2 [64.5 80.0]** | **68.4 [56.1 80.6]** | **76.1 [68.0 84.2]** | **73.2 [63.0 83.3]** | **70.3 [60.9 79.7]** | **0.72 [0.65 0.80]** |
| *Iowa City* | 60.6 [40.6 80.7] | 36.1 [14.9 57.3] | 100 [100 100] | 100 [100 100] | 51.1 [30.9 71.3] | 0.59 [0.46 0.73] |
| *Medellín* | 68.7 [55.8 81.6] | 73.5 [56.5 90.4] | 63.9 [45.8 81.9] | 67.5 [50.2 84.7] | 69.5 [55.1 84.0] | 0.69 [0.56 0.82] |
| *San Diego* | 77.7 [61.0 94.3] | 74.5 [48.5 100] | 81.6 [59.0 104] | 78.8 [53.4 104] | 74.2 [53.4 104] | 0.78 [0.60 0.95] |
| *Turku* | 82.2 [67.8 96.6] | 82.1 [62.2 103] | 82.5 [62.1 103] | 82.6 [62.5 103] | 81.3 [64.2 98.5] | 0.82 [0.66 0.98] |

Since LR on harmonized features with univariate feature selection was the model with the most promising validation results, it was selected for the final classification evaluation (Table IV). An average accuracy of 72.2% was achieved for all the centers together, balanced by average accuracies of 60.6%, 68.7%, 77.7%, and 82.2%, for Iowa City, Medellín, San Diego, and Turku datasets, respectively. Worthy of note that the Iowa City dataset performed worse compared to all other datasets. A possible explanation is that it is the smallest dataset resulting in a limited number of subjects for both training and testing.

Some limitations of our work include a limited amount of test data on model selection in the nested CV approach, which can be explored in future work.

IV. CONCLUSION

Our results show that harmonization improves global classification performance. Our end-to-end framework can support the clinical diagnosis of PD. Our work shows the importance of harmonization in multi-center settings while being limited by the test of only one harmonization technique we consider relevant. Future work might include testing different approaches for harmonization.

ACKNOWLEDGMENT

We would like to thank all the researchers and staff at the University of California San Diego, the University of Turku and the Turku University Hospital, the University of Iowa, and the University of Antioquia. We especially express our gratitude to Dr. A. P. Rockhill, Dr. H. Railo, and Dr. N. S. Narayanan for making their datasets publicly available.